\documentstyle [12pt,epsfig,aaspp4] {article}

\newcommand{\be}{\begin{eqnarray}}
\newcommand{\ee}{\end{eqnarray}}
\def\lsim{\mathrel{\rlap{\lower3pt\hbox{\hskip1pt$\sim$}}
    \raise1pt\hbox{$<$}}} 
\def\gsim{\mathrel{\rlap{\lower3pt\hbox{\hskip1pt$\sim$}}
    \raise1pt\hbox{$>$}}} 
\newcommand{\msun}{M_\odot}
\newcommand{\rsun}{R_\odot}

\newcommand{\He}{\rm He}

\def\bi{\begin{enumerate}}
\def\ei{\end{enumerate}}

\def\astrobjV404Cyg{V404 Cyg }
\def\astrobjCygX1{Cyg X-1}

\def\msun{M_\odot}

\def\kpc{kpc}

\lefthead{Lee, Brown, \& Wijers}
\righthead{Soft X-ray Transients, Gamma-Ray Burst, and Hypernova}

\begin{document}

\title{Discovery of a Black-hole Mass - Period Correlation in 
       Soft X-ray Transients
       and its Implication for Gamma-Ray Burst and Hypernova Mechanisms}
\author{C.-H. Lee$^{a,b,c}$, G.E. Brown$^{c}$, and R.A.M.J. Wijers$^{c}$}
\affil{$a)$ School of Physics, Seoul National University, Seoul 151-742, Korea\\
        $b)$ School of Physics,
       Korea Institute for Advanced Study, Seoul 130-012, Korea\\
        $c)$ Department of Physics \& Astronomy,
        State University of New York,
        Stony Brook, New York 11794, USA\\
E-mails: chlee@kias.re.kr, popenoe@nuclear.physics.sunysb.edu,
rwijers@mail.astro.sunysb.edu}

%
\begin{abstract}
We investigate the soft X-ray transients
with black hole primaries which may have been the sources of
gamma-ray bursts and hypernovae earlier in their evolution. 
For systems with evolved donors we are able to reconstruct
the pre-explosion periods and find that the black-hole mass
increases with the orbital period of the binary.
This correlation can be understood in
terms of angular-momentum support in the helium star progenitor 
of the black hole, if the systems
with shorter periods had more rapidly rotating primaries prior to their
explosion; centrifugal support will then prevent more of its
mass from collapsing into the black hole on a dynamical time.
This trend of more rapidly
rotating stars in closer binaries is usual in close binaries, and in the 
present case can be understood in terms of spin-up during spiral in
and subsequent tidal coupling. We investigate the relation quantitatively
and obtain reasonable agreement with the observed mass-period correlation.
An important ingredient is the fact that the rapidly rotating new black
hole powers both a GRB and the hypernova explosion of the remaining
envelope, so that the material initially prevented from falling into
the black hole will be expelled rather than accreted.

For systems in which the donor is now, and will remain, in main
sequence we cannot reconstruct the pre-explosion period in detail,
because some of their history has been erased by angular momentum
loss through magnetic braking and gravitational waves. We can, however,
show that their periods at the time of black hole formation were most
likely 0.4--0.7 day, somewhat greater than their present periods.
Furthermore, their black holes would have been expected to accrete
$\sim 1\msun$ of material from the donor during their previous
evolution. Comparison with predictions suggests that little mass
will be lost in the explosion for the relatively high pre-explosion
periods of these binaries.

A natural consequence of the He star rotation is that black holes formed
in the shorter period (before explosion)
soft X-ray transients acquire significant Kerr parameters. This
makes them good sources of power for gamma-ray bursts and hypernovae,
via the Blandford-Znajek mechanism, and thus supports our model for the
origin of gamma-ray bursts in soft X-ray transients.

\end{abstract}



%

\newpage
\section{Introduction}

Recent observations strongly suggest a connection between gamma-ray
bursts and supernovae, with indication that the supernovae in 
question are especially energetic and of type Ib/c, i.e., core
collapses of massive stars which have lost their hydrogen envelope
(see Van Paradijs, Kouveliotou, and Wijers 2000, and references therein).
This supports suggestions by Woosley (1993) and Paczy\'nski (1998)
for the origin of gamma-ray bursts in stellar core collapses. The 
hydrodynamics of a jet escaping from a star and causing its explosion
was explored in detail by MacFadyen and Woosley (1999), who showed that
contrary to accepted wisdom, a fairly baryon-free, ultra-relativistic 
jet could plow through the collapsing star and emerge with large Lorentz
factors. The powering of the outflow by coupling of high magnetic fields
to the rotation of the black hole (Blandford \& Znajek 1977), first
suggested by Paczy\'nski (1998) in the context of gamma-ray bursts, was
worked out in detail by Van Putten (1999, 2001).
Li has also discussed the deposition of energy from a black hole
into the accretion disk in a recent series of papers (e.g., Li 2000a).

Building on these thoughts, we have modeled both the powering of a
gamma-ray burst by black-hole rotation and the stellar evolution pathways
that set up favorable conditions for that mechanism (Brown et al. 2000).
An essential ingredient in this model is a rapidly rotating black hole,
and it is this aspect that we focus on in the present paper. A single star
initially in uniform rotation will tend to develop a differential rotation
because the core contracts strongly during evolution, and angular-momentum
conservation will therefore increase its angular velocity. However,
given enough time viscous stresses will even out these differences, and
thus the net result is a loss of angular momentum of the innermost regions
of the star. Spruit \& Phinney (1998) argued that magnetic-field mediated
coupling is strong enough in single stars during the giant phase to
make the cores very slow; so slow, in fact, that they required an asymmetric
kick in the birth of pulsars to get their spin frequencies up to 
observed values. Livio \& Pringle (1998) subsequently used observations of
novae to argue for a weaker coupling, but still their coupling strength 
would lead to spin energies of black holes that are negligible as 
power sources for gamma-ray bursts.

However, as suggested by MacFadyen \& Woosley (1999), a massive star in
a close binary will spin faster for a number of reasons: first, when the
hydrogen envelope is lifted off by spiral-in, it will cease to serve as a 
sink of angular momentum for the core. Second, the tidal friction
concomitant of the spiral-in process will spin up the inner region,
giving it a larger angular momentum than the same region in a single star
(Rasio and Livio 1996).
Third, tidal coupling in the close binary will tend to bring the primary
into corotation with the orbital period. 
This latter process is not very efficient in
the short post spiral-in life of the binaries we consider, but its effect
does probably matter to the outer layers of the helium star, which can be
important for our work. With its more rapid rotation, the helium star
then forms a black hole with a large Kerr parameter, which immediately
after its formation (in a few seconds) begins to input power into its
surroundings at a very high rate. This, then, powers both a gamma-ray burst
(e.g., Brown et~al.\ 2000) and the expulsion of the material that was
centrifugally prevented from falling into the black hole. In fact,
Van Putten (1999, 2001) estimates that the power input into that material
exceeds that into the GRB and Li (2000b) also finds that more energy
can be extracted by the disk than by the GRB. 
It should be noted that an initially less
rapidly rotating black hole could be spun up by disk accretion quite
rapidly, and start a similar process after some accretion has taken
place (MacFadyen \& Woosley 1999, Brown et~al.\ 2000). Some implications
of such more complicated sequences of events are discussed by Lee, Lee,
\& Van Putten (2001).

In section~\ref{porb} we present the data on known soft X-ray transients,
showing the relation between present orbital period and black-hole mass.
Since theory predicts a relation between pre-explosion orbital period
and black-hole mass, we then consider carefully the pre- and post-explosion
evolution of the systems (section~\ref{evol}), and use this to
reconstruct the pre-explosion orbit for as many 
systems as possible (section~\ref{preexp}).
Then we develop our
model for the mass and spin of black holes in soft X-ray Transients (SXTs), 
and use it to explain
the mass-period correlation (section~\ref{rotate}). 
We summarize our conclusions in section~\ref{concl}.

\section{An empirical correlation between orbital period and black-hole mass
         \label{porb}}

\begin{figure}
\centerline{\epsfig{file=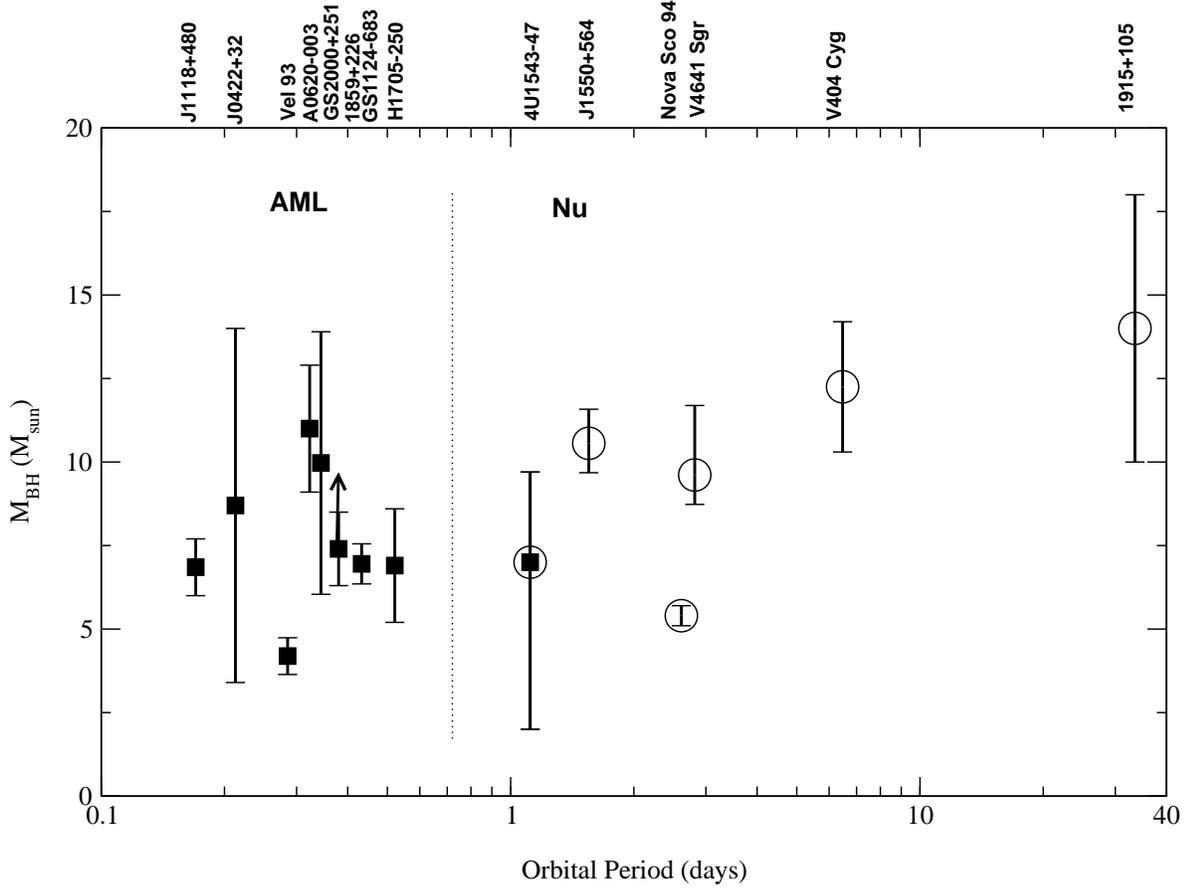,height=5in}}
\caption{Black hole mass as a function of present orbital period of 14 SXTs. 
Note that the orbital period is on a logarithmic scale. 
SXTs with subgiant or giant companions
are indicated by big open circles (denoted as ``Nu" for nuclear evolution). 
Filled squares indicate 
SXTs with main-sequence companions 
(denoted as ``AML" for angular momentum loss).
The vertical dotted line is drawn to indicate the possible
existence of different classes according to evolutionary path of the
binary, as discussed in section~\ref{evol}.
4U 1543$-$47 is marked with both symbols, since we believe it to
be right on the borderline between main-sequence and evolved; for the
purpose of modeling, it can be treated as evolved.
}
\label{FIG1}
\end{figure}

We have collected data from the literature on black-hole binaries in
our Galaxy. In table~1 we collect data of those in which the mass function
is known, and some manner of mass estimate for both the black hole and 
companion can be given. In table~2 we list the properties of two key systems
in more detail. In Fig.~\ref{FIG1} we show the masses of the black
holes as a function of orbital period. While the ranges of black-hole
masses for main-sequence and evolved systems overlap, the latter tend to
have higher masses; the exception is Nova Scorpii 1994, which we shall
see later is a natural but rare case of the general evolution scenario
that we describe in this paper. In Fig.~\ref{FIG2}, we show the 
donor masses as a function of orbital period. They show a more obvious
trend of more massive donors in evolved systems. As we shall see, this
is a natural consequence of the fact that only evolved systems can come
into Roche contact in wide binaries, and more massive donors are more
likely to come into contact via nuclear evolution. (The various curves
are explained in section~\ref{evol}.)

In the following sections, we shall argue that the correlation between
black-hole mass and period also has physical meaning: the shorter the
orbital period, the more rapidly rotating the helium star progenitor 
to the black hole. Rapid rotation centrifugally prevents some
fraction of the helium 
star to collapse into a black hole, resulting in a smaller black hole mass.
The reason why the correlation in Fig.~\ref{FIG1} is weak is that evolution
of the binary since formation of the black hole has washed out the 
relation. Properly, we should consider the correlation between pre-explosion
orbital period and post-explosion black hole mass. Much of our work
presented here is concerned with understanding the evolution of these
binaries, and using this knowledge to find the systems for which we can
reconstruct those parameters. Using that subset, we show much better
agreement between our model predictions and the observed
relation between reconstructed period and mass; this supports our
evolutionary model and has ramifications for the origin of gamma-ray bursts.

\section{The evolution of soft X-ray transients\label{evol}}

   \subsection{Prior to the formation of the black hole\label{evol.gen}}

Following on the work of Brown, Weingartner, \& Wijers (1996), who
showed the importance of mass loss of helium stars in binaries in
determining the final outcome of binary evolution,
Brown, Lee, \& Bethe (1999), Wellstein \& Langer (1999),
and Brown et al. (2001a) showed that massive
helium stars could evolve into high-mass black holes only if they were
covered with hydrogen during most of their helium core burning era
(Case C mass transfer in binaries). In case A
or B mass transfer in binaries (Roche Lobe overflow in main-sequence
or red-giant stage) the Fe core that was left was too low in mass
to go into a high-mass black hole.
Brown et~al.\ (2001a) showed that high-mass black holes
could be formed only if the mass was taken off of the black hole
progenitor after helium core burning was finished; i.e. Case C mass
transfer. Brown, Lee, \& Tauris (2001c) showed that with the Schaller et al.
evolution, this could happen only in the neighborhood of ZAMS mass 
$20\msun$, definitely not at $25\msun$ and higher: Because of the wind
losses, the mass transfer would begin as Roche lobe overflow only in Case B 
mass transfer, with these higher main sequence masses, because the giant
radii of these stars exceed their radii in the supergiant phase.
Below we shall see that the high black-hole masses
in a few binaries require us to extend the range upward to about 30\,$M_\odot$.

A major uncertainty in the evolution of all compact X-ray binaries is
the phase of spiral-in that occurred in their evolution: these binaries
are initially very wide, and when the primary fills its Roche lobe and
transfers mass to the secondary, the mass transfer leads to instability,
resulting in the secondary plunging into the primary's envelope. Next,
dissipation of orbital energy of the secondary causes the primary's envelope
to be ejected, and the orbit to shrink. Following the original work by
Webbink (1984), Brown, Lee, \& Tauris (2001c)
write the standard formula for common envelope evolution as
    \be
    \frac{G M_pM_e}{\lambda R} 
    = \frac{G M_pM_e}{\lambda r_L a_i} 
    = \alpha_{ce} 
     \left(\frac{G M_{\He} M_d}{2 a_f}-\frac{G M_p M_d}{2a_i}\right)
    \label{eq4}
    \ee
where $M_p$ is the total mass of the BH progenitor star just before
the common envelope forms, $M_e$ is the mass of its hydrogen envelope,
$M_{\He}$ is the mass of its core, $a_{\rm i}$ and $a_{\rm f}$ are the
initial and final separation, before and after the common envelope,
respectively, and $r_L\equiv R_L/a$ 
is the dimensionless Roche-lobe radius. This
equation essentially relates the loss of orbital energy of the secondary
to the binding energy of the ejected envelope. The parameter $\lambda$
is a shape parameter for the density profile of the envelope. It can vary
greatly between stars (Tauris \& Dewi 2001), but for the extended, deeply
convective giants we deal with in case C mass transfer it is always
close to 7/6. (See also Appendix C of Brown et al. 2001b.)  The parameter
$\alpha_{ce}$ accounts for the efficiency with which orbital energy is
used to expel the envelope, and may also account for some other effects
such as extra energy sources and the possibility that each mass element
of the envelope receives more than the minimum energy needed to escape
(see, e.g., Bhattacharya \& Van den Heuvel 1991 and references therein).

Given the parameters of the system
at first Roche contact, when spiral-in starts, the final separation is
determined by the product of $\lambda$ and $\alpha_{ce}$, 
the efficiency of the energy conversion. In general, these parameters are
only the simplest recipe prescription for the complex hydrodynamical 
interaction during spiral-in. While we therefore cannot predict the value
of $\lambda\alpha_{ce}$ from first principles, we can try to find its
value from constraints in some systems, and then assume it is the same for all
similar systems. 
Brown, Lee, \& Tauris found a great regularity in the evolution of SXTs
with main-sequence companions, all but one of which are K or M stars,
which constrained the efficiency $\lambda\alpha_{ce}$ to be
0.2--0.5. However, these authors did not include mass loss
in the explosion, which we shall do here in our evolution of SXTs with
evolved companions. Since mass loss substantially
widens the orbits, including it the common envelope evolution must
bring the (pre-explosion) $a_f$ to a smaller value: if 
$M_{post}$ is the black hole plus companion mass, and $\Delta M$ the mass
lost in the formation of the black hole, we have
  \be
   a_{f,post}= a_{f,pre}(1+\Delta M/M_{post}).
  \ee 
(after the orbit has been re-circularized).
Therefore, there has been an extra widening since the explosion by a factor
of up to about 1.5. 
We found $\lambda\alpha_{ce}$ to be in the lower part of the interval
found by Brown, Lee, \& Tauris (2001c), $\lambda\alpha_{ce}\sim 0.2$.

We can achieve a more precise `calibration' of the value of
$\lambda\alpha_{ce}$ if we manage to find some systems in which we can
estimate both the initial and final separation.  To estimate the initial
separation (at the onset of spiral-in) we need to know the mass and
radius of the black-hole progenitor and combine this with the Roche-lobe
filling condition.  The helium star progenitors in at least three
of the evolved binaries seem to be too massive for the $20-23\msun$
ZAMS progenitors used by Brown, Lee, \& Tauris (2001c): the black
hole in V404 Cyg is probably at least $10\msun$ (Shahbaz et al. 1994,
Shahbaz et al. 1996, Bailyn et al. 1998), and the black hole in Nova
Scorpii is of mass $\sim 5.4\pm 0.3\msun$ (Beer \& Podsiadlowski 2001)
and the mass loss in black hole formation is $\gsim 5\msun$ (Nelemans,
Tauris, Van den Heuvel 1999) so that the progenitor of the helium star
must have been
$\sim 11\msun$.  {}From Table~\ref{tab2}, the black hole in V4641 Sgr is
of mass $9.61^{+2.08}_{-0.88}\msun$ (Orosz et al. 2001).  The tentative
conclusion from the above is that at least these binaries with evolved
companions seem to have come from helium cores of $\sim 11\msun$, or ZAMS
mass $\sim 30\msun$.
With high wind mass loss rates as proposed by Schaller et~al.\ (1992),
such massive stars have larger radii as giants than as supergiants,
thus making case C mass transfer impossible.  However, since radii
and mass loss rates of evolved stars are very uncertain, we take the
view that the need for $\sim 11\msun$ helium cores implies that their
progenitors, $30\msun$ main-sequence stars, do expand enough to allow
case C mass transfer.

In Fig.~\ref{FIG3}, we summarize the radius at the end
of the giant branch as a function of ZAMS mass (Schaller et~al.\ 1992).
The ZAMS mass dependence of this final giant radius is adequately represented
by a linear function in the region of $20-40\msun$. 
We assume that the radial expansion during the helium burning can be
scaled to the case of a $20\msun$ star using this linear relation as follows:
   \be
   R (M;t) = \left( - \frac{842\rsun-750\rsun}{842\rsun}
      \frac{(M-20\msun)}{20\msun} +1\right) R(20\msun;t).
   \ee
Further, we took the mass loss rate of $20\msun$ as standard, and
scaled the mass loss rate in proportion to the ZAMS mass.  The allowed
range of Case C mass transfer with ZAMS mass $20\msun$ is $971\rsun <
R < 1185\rsun$, that of Schaller et~al.\ (1992). In Fig.~\ref{FIG4} are
given the possible initial orbital separations for Case C mass transfer
for the $1.91\msun$ companion appropriate for Nova Scorpii 
(see section~\ref{preexp}) and for the $6.53\msun$ companion appropriate for
V4641\,Sgr (Orosz et~al.\ 2001).

Now if we look at eq.~(\ref{eq4}), we see that $a_f$ scales almost
linearly with the donor (companion) mass $M_d$. The envelope mass $M_e$
is roughly $0.7 M_{giant}$ (Bethe \& Brown 1998) and we use
    \be
    M_{He}=0.08 (M_{giant}/\msun)^{1.45} \msun
    \ee
so that
    \be
    a_f \propto \frac{M_d}{\msun} \left(\frac{M_{giant}}{\msun}\right)^{-0.55}
    a_i
   \label{eqaf}
   \ee
assuming $\lambda\alpha_{ce}$ to be constant and with neglect of the small
term in $a_i^{-1}$ in the r.h.s of eq.~(\ref{eq4}).  {}From our curves,
Fig.~\ref{FIG4}, we see that the 20\% possible variation in $a_i$ results
in the same percentage variation in $a_f$. Because the
actual ZAMS mass can be anywhere in the range
$20-30\msun$ there can be an additional $\sim 25\%$ variation in $a_f$
with giant mass, as compared with the linear dependence on $M_d$. In
view of the modest size of these variations at a given donor mass, we make the
approximation in the rest of the paper that the pre-explosion orbital
separation depends only on $M_d$, and scales linearly with $M_d$.
This simple scaling and the modest amount of scatter around it are
partly the result of the weak dependences on initial parameters in 
eq.~(\ref{eqaf}), but chiefly the result of the fact that our
model uses case C mass transfer. This constrains the Roche contact to
first occur when the radius of the star is in a very narrow range, 
between the maximum radius in the giant phase and the maximum radius
in the supergiant phase.

To complete our calibration of the spiral-in efficiency, we need to
find systems in which we can also estimate the orbital separation
just after spiral-in well. This is complicated by the fact that mass
transfer has taken place since the spiral-in. Most SXTs have small
mass ratios, and for such small mass ratios the orbital separation
is fairly sensitive to the amount of mass transferred, making it
hard to derive the post-spiral-in separation from the present one.
The exception is V4641\,Sgr, in which the present mass ratio is close
to 1. Since the initial mass ratio could not have been significantly
greater than 1 (since that would result in unstable mass transfer), and
furthermore the orbital period changes very little with mass transfer
for nearly equal masses, we can fairly approximate the post-spiral-in
separation by the present one. In Fig.~\ref{FIG5}, we show the 
predicted ranges of post-spiral-in orbital periods for different values
of $\lambda\alpha_{ce}$. Clearly, a value quite close to 0.2 is
indicated. 
For 4U\,1543$-$47 (IL Lupi), we find that it is near the boundary
between evolved and main-sequence evolution. To place it there,
as discussed in section~\ref{preexp},
we find from the reconstructed orbital
period in Fig.~\ref{FIG6} that $\lambda\alpha_{ce}\sim 0.2$ is also
consistent with the properties of this system.

In short, the general properties of SXTs and the specific cases of
V4641\,Sgr and IL Lupi favor $\lambda\alpha_{ce}\sim 0.2$, which we 
therefore adopt as a general efficiency 
for the evolution of other transient sources. This then makes it possible
to make quite specific predictions for the prior evolution of many of the
other SXTs.

   \subsection{Expected regularities}

{}From the above theory, certain regularities follow for the system
behavior as a function of companion mass. First, the binding energy
relation for spiral-in (eq.~(\ref{eq4})) shows that very nearly $a_f\propto
M_d$, with not much variation due to other aspects of the systems
(see previous section). Furthermore, the relation between Roche lobe
radius and donor mass when $M_d\ll M_{BH}$ implies that $R_L/a_f\propto
M_d^{1/3}$ (e.g., Eggleton 1983). As a result, the Roche lobe radius of
the donor just after spiral-in will scale with donor mass as $R_L\propto
M_d^{4/3}$. On the other hand, the donor radius itself depends on
its mass only as $R_d\propto M_d^{0.8}$.  Therefore, a low-mass donor
overfills its Roche lobe immediately after spiral-in. In the donor mass
range we consider ($M_d\gsim0.7\msun$) it does not overfill its Roche lobe
by much, so we assume that the system adjusts itself quickly by transfer
of a small amount of mass to the He star, which widens the orbit until
the donor fills its Roche lobe exactly.
Above this minimum mass, there will be a range
of donor masses that are close enough to filling their Roche lobes
after spiral-in that they will be tidally locked and will come into
contact via angular momentum loss (AML). 
Above this, there will be a range of mixed evolution,
where both AML and nuclear evolution (Nu) play a role. Finally, for the most
massive donors, $M_d>2\msun$, the post-spiral-in orbits will be too
wide for AML to shrink them much, so mass transfer will be initiated
only via nuclear expansion of the donor.  Of course, the ranges of case
C radii of stars and variations of primary masses will ensure that the
boundaries between these regions are not sharp: near the boundaries the
fate of the system depends on its precise initial parameters.

\section{Reconstructing the pre-explosion orbits\label{preexp}}

{\it Nova Sco 94 (GRO\,J1655$-$40):\/}
The most extensive evolutionary studies have been made for Nova Scorpii.
Starting from the work of Reg\H{o}s, Tout \& Wickramasinghe (1998) who
make the case that the companion is in late main sequence evolution,
Beer \& Podsiadlowski (2001) carry out extensive numerical calculations of the
evolution, starting with a pre-explosion mass of $2.5\msun$ and separation
of $\sim 6\rsun$. More schematically we arrived at a pre-explosion
mass of $1.91\msun$ and separation of $5.33\rsun$.
We consequently have an 0.4 day pre-explosion period. With $\sim 6\msun$
mass loss in the explosion (Nelemans et al. 1999), nearly half the system
mass, the binary period increases to 1.5 day, well beyond the period gap. 
This is also the period required if the common-envelope efficiency
in this binary was again 0.2 (Fig.~\ref{FIGSco}).
This explains why Nova Sco is the only system with a giant donor and a
black-hole mass in the lower end of the range: its evolution really places
it among the narrow-orbit systems. Generally, the mass loss during explosion
is mild, and does not change which category a system belongs to. But in
those exceptional cases where the mass loss comes close to half the
total mass, the orbit widens very much and converts an AML system to a
nuclear-evolution system.
We shall discuss in section~\ref{concl} that help in expelling the mass
may come from early onset of the GRB mechanism.
After explosion the binary evolves to its 
present period by nearly conservative mass transfer. Our estimate is
that $0.41\msun$ is transferred from the donor to the black hole.
Brown et~al.\ (1999) first made the case that Nova Scorpii was the
relic of a GRB.

{\it V4641\,Sgr:\/}
As we discussed in section~\ref{evol}, this system is our calibrator 
for the spiral-in efficiency, and we assume that its present state is
very close to the one immediately following spiral-in.

{\it GRS 1915$+$105:\/}
Recently Greiner et al. (2001) have determined the period and black hole
mass of GRS 1915$+$105 to be 33.5 day and 14$\pm 4\msun$. Interestingly,
we can evolve a system with properties very close to this by simply starting
from V4641\,Sgr and following its future evolution with conservative
mass transfer ($P_{orb}\propto\mu^3$, where $\mu$ is the reduced mass);
allowing for $4.6\msun$ to
be transferred from the donor to the black hole, we have
   \be
   P_{1915}=\left(\frac{9.61\times 6.53}{14.21\times 1.93}\right)^3
   P_{4641} = 33.7 \; {\rm day} .
   \ee
This would give a companion mass of $1.93\msun$, as compared with the
Greiner et al. (2001) mass of $M_d=1.2\pm 0.2\msun$.
However, the mass transfer cannot be completely conservative because of loss
by jets, etc., as evidenced by the microquasar character of this object. 
Furthermore the above $M_d$ is viewed as a lower limit by Greiner
et al. because the donor is being cooled by rapid mass loss, but its mass is
estimated by comparison with non-interacting stars. We thus believe our
evolution to be reasonable. We position the pre-explosion period and black hole
mass of GRS 1915$+$105 at the same point as V4641 Sgr. Since mass transfer
and widening of the orbit always occur together, the effect of this 
post-explosion evolution is to
introduce a weak secondary correlation between orbital period and companion
mass in the long-period regime, where such a correlation is not expected to
arise from the pre-explosion evolution.

{\it IL Lupi:\/}
Recently over-abundances of Mg in the companion star of IL Lupi have been
observed (Orosz 2002). In analogy with the case of the overabundances
in Nova Scorpii (Israelian et~al.\ 1999, Brown et~al.\ 2000), this
indicates that there was an explosion at the time of black hole
formation in this system, in which some of the material ejected from
the core of the helium-star progenitor to the black hole ended up on
the companion.  Based on these observations and our given efficiency
$\lambda\alpha_{ce}=0.2$, one can start with $11\msun$ He star and
$1.7\msun$ companion as a possible progenitor of IL Lupi.  {}From the lower
boundary of the curve with $\lambda\alpha_{ce}=0.2$ in  Fig.~\ref{FIG6},
the period would be 0.5 days.  By losing $4.2\msun$ during the explosion,
the binary orbit would be widened to 1.12 day.  The period had to be
shortened to 0.8 day by magnetic braking and gravitation wave radiation
before the mass transfer started.  Conservative transfer of $0.23\msun$
from the companion to the black hole would bring the period from 0.8
day to the present 1.1164 day.

{\it V404\,Cyg:\/}
The black hole in V404 Cyg appears to be somewhat more massive than in IL Lupi,
so we begin with a similar mass companion, but a $10\msun$ black hole, which
would have a period of 0.63 day. Again, we neglect mass loss in the explosion,
although a small correction for this might be made later. Conservative
transfer of $1\msun$ from the donor to the
black hole then brings the period to
   \be
   0.63 \; {\rm day} \left(\frac{1.7\msun\times 10\msun}{0.7\msun\times 11\msun}\right)^3 
   = 6.7\;  {\rm day}
  \ee
close to the present $6.47$ day period. Here we take $11\msun$ and $0.7\msun$
as current masses in V404 Cyg (Orosz 2002).
The black hole in V404 Cyg seems to be somewhat more massive than the others
in the transient sources, with the exception of that in GRS 1915$+$105. 
In both cases we achieve the relatively high black hole masses
and periods by substantial accretion onto the black hole.

{\it GRO J1550$-$564:\/}
The high mass black hole in J1550$-$564, $10.56\msun$ (Orosz et~al.\
2002), is slightly 
less massive than the assumed black hole mass of V404 Cyg, 
and the companion is more massive than V404 Cyg with short
period, 1.552 days. So, we start from the same initial
conditions just derived for V404\,Cyg (Fig.~\ref{FIG11}), and end up
with the present system via simple conservative mass transfer.

{\it Cygnus X-1:\/}
Cyg\,X-1 is usually not considered to have come from the same evolutionary
path as the SXTs, since it is a persistent X-ray source with a much more
massive donor. But with the discovery of objects with relatively massive
donors in the SXT category, such as V4641\,Sgr, it is worth considering the
implications of our model for it.
Cyg X-1\,has been shown to have an appreciable system velocity
(Kaper et~al.\ 1999) although it may be
only $1/3$ the $50$ km s$^{-1}$ given there, depending on the O-star
association (L. Kaper, private communication). The evolution of Cyg X-1
may have been similar to that of the transient sources, the difference
being in the copious mass loss from the companion O9I star, causing the
black hole to accrete and emit X rays
continuously. If we scale to Nova Scorpii to obtain the
initial binary separation, we find
   \be
   a_f =\frac{17.8\msun}{1.91\msun}\times 5.33 \; \rsun =50\; \rsun
   \ee
somewhat larger than the present binary separation of $40\rsun$. (We would
obtain $38\rsun$ if we scaled from the Beer \& Podsiadlowski (2001) companion mass
of $2.5\msun$ for Nova Scorpii.)
Given uncertainties in the mass measurements, we believe it possible for
Cyg X-1 to be accommodated in this scheme. Some sort of common envelope
envelope evolution seems to be necessary to narrow the orbit in the
evolution involving the necessarily
very massive progenitor stars
(Brown et al. 2001a).

\subsection{Problems with the close (AML) systems\label{preexp.aml}}

\begin{figure}
\centerline{\epsfig{file=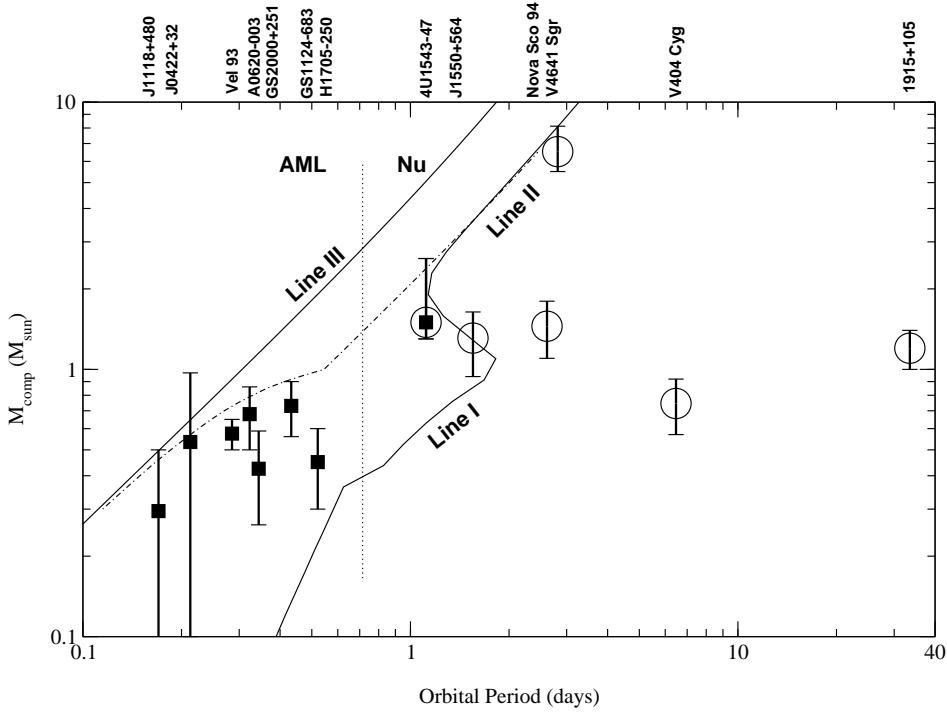,height=4.0in}}
\caption[]{Companion mass as a function of present orbital period of 13 SXTs. 
XTE 1859$+$226 is not included because the companion mass
is not well determined (Filippenko \& Chornock 2001).
Symbols of SXTs are the same as in Fig.~\ref{FIG1}. Line III indicates the
orbital period for which a companion of that mass fills its Roche lobe on
the ZAMS. No system can exist above and to the left of this line for 
a significant duration. Lines I and II are the upper period limit for systems
that can come into contact while the donor is on the main sequence. For
high masses (line II) this limit is set by the period where the evolution
time of the companion is too short to allow the orbit to shrink significantly
before it leaves the main sequence.  For low masses
(line I), where the donor never evolves off the main sequence within a 
Hubble time, the limit is set by period for which the shrinking time scale
of the orbit equals the Hubble time. The dot-dashed line indicates the point
where a system that starts its life on lines I/II comes into Roche contact.
For very low masses, this equals line III, because the donor never 
moves significantly away from its ZAMS radius, whereas for very high masses
it equals line II, because the orbit cannot shrink before the companion
evolves off the main sequence. At intermediate masses, the companion expands
somewhat while the orbit shrinks, and fills its Roche lobe at a larger
period than line III. Systems that become SXTs with main-sequence donors
within a Hubble time must start between line III and line I/II. At the start
of mass transfer, they must lie in the narrow strip between line III and
the dot-dashed line.
}
\label{FIG2}
\end{figure}

\begin{figure}
\centerline{\epsfig{file=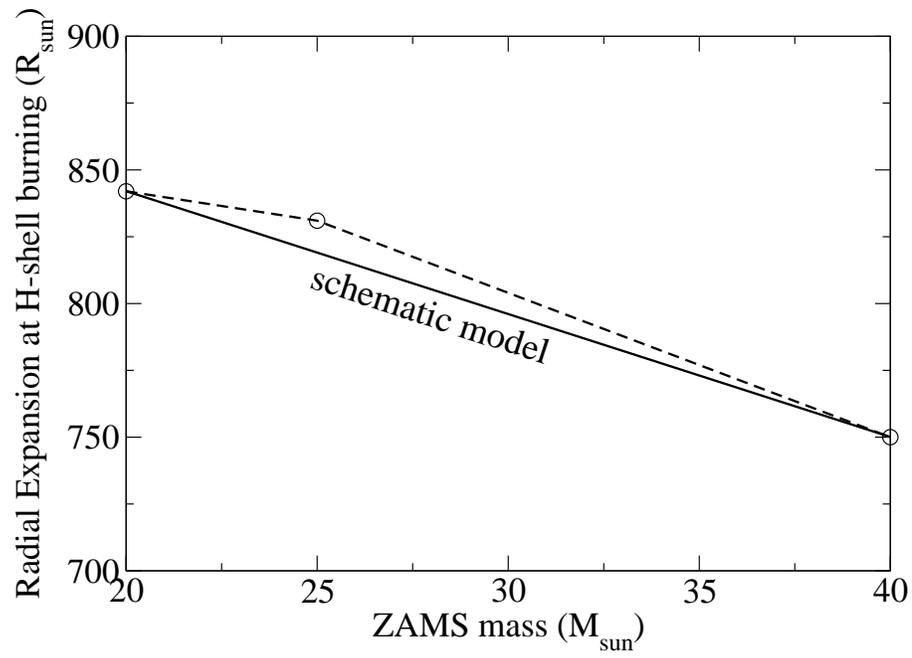,height=4in}}
\vskip 5mm
\caption{Radial expansion at the end of the giant branch. The data taken from
Schaller et al. (1992). Radial expansions are
$842\rsun$, $831\rsun$, and $750\rsun$ for
$20\msun$, $25\msun$, and $40\msun$, respectively.
The radial expansion shows an almost linear dependence
on the ZAMS mass.}
\label{FIG3}
\end{figure}

\begin{figure}
\vskip 10mm
\centerline{\epsfig{file=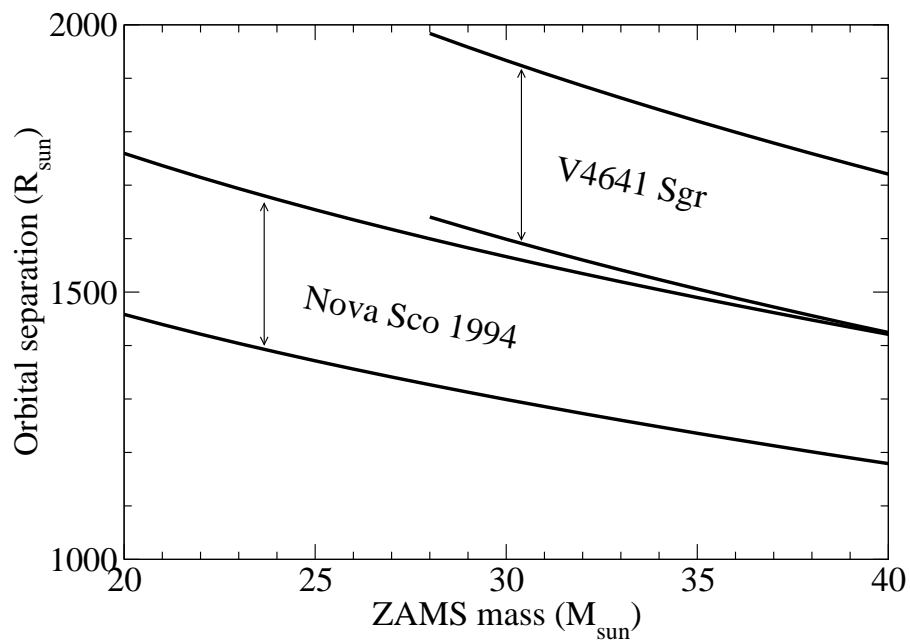,height=4.0in}}
\vskip 5mm
\caption{Possible initial orbital separations for Case C mass transfer
with $1.91\msun$ (Nova Sco) and $6.53\msun$ (V4641 Sgr)
main-sequence companions. For Nova Sco, the estimated companion mass 
$1.91\msun$ during the common envelope evolution
is used as discussed in section~\ref{evol}.
The x-axis is the ZAMS mass of the black hole progenitor. 
}
\label{FIG4}
\end{figure}

\begin{figure}
\centerline{\epsfig{file=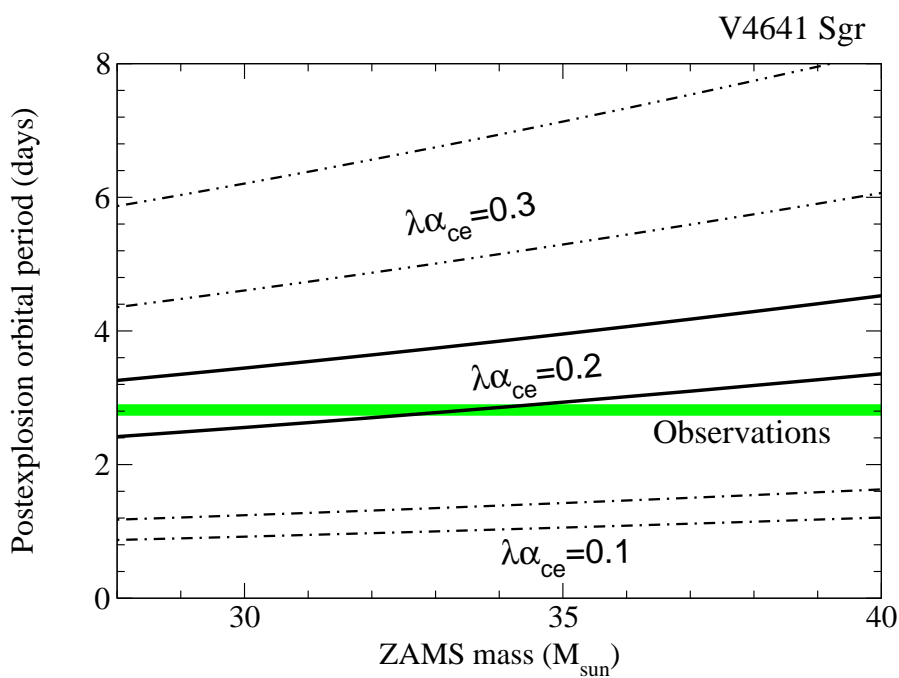,height=4.0in}}
\caption{Post-explosion orbital periods of V4641 Sgr
after BH formation for various common envelope efficiencies.
The x-axis is the ZAMS masses of black hole progenitor. 
The width of each band
is determined by the initial band of possible Case C
mass transfer given in Fig.~\ref{FIG4}. 
$\lambda\alpha_{ce} \sim 0.2$ is consistent with current observations. 
}
\label{FIG5}
\end{figure}

\begin{figure}
\begin{center}
\epsfig{file=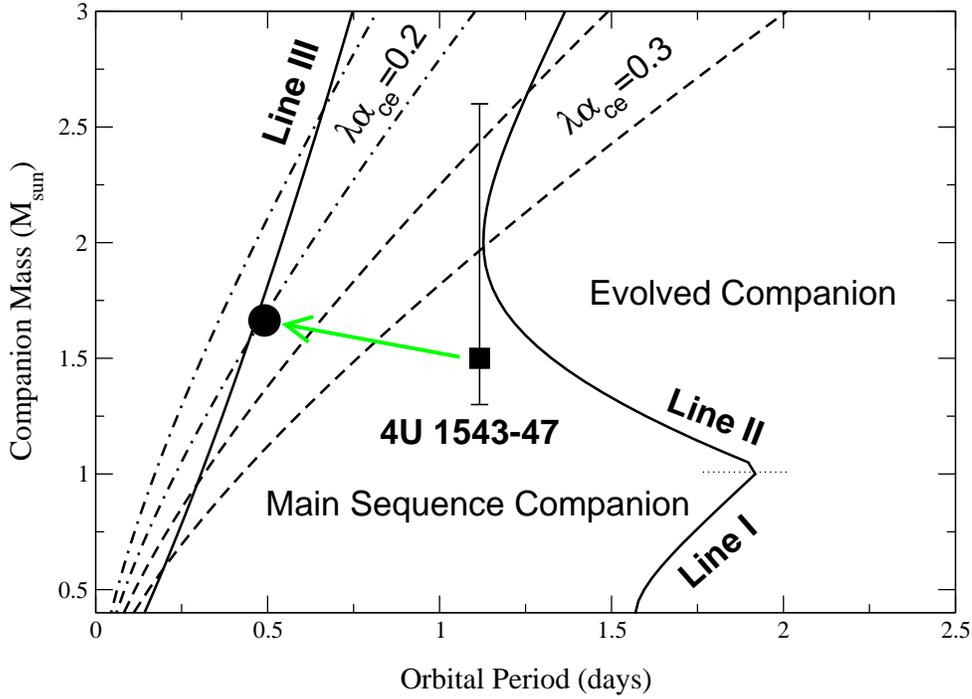,height=4.5in}
\end{center}
\caption{
  An enlargement of the central region of Fig.~\ref{FIG2}, indicating
  the present location of IL Lupi. If we adopt the same value
  $\lambda\alpha_{ce}=0.2$ for the common-envelope parameter, we see
  that the reconstructed post-spiral-in period places the system
  right on the main-sequence line, implying a fairly tight constraint
  on the initial parameters of this system.
  (We took the ZAMS mass of the black hole progenitor to
  be $30\msun$, which corresponds to $M_p\sim 25\msun$ in the
  beginning of case~{C} mass transfer, and $M_{He} = 11\msun$.)
  }
\label{FIG6}
\end{figure}

\begin{figure}
\centerline{\epsfig{file=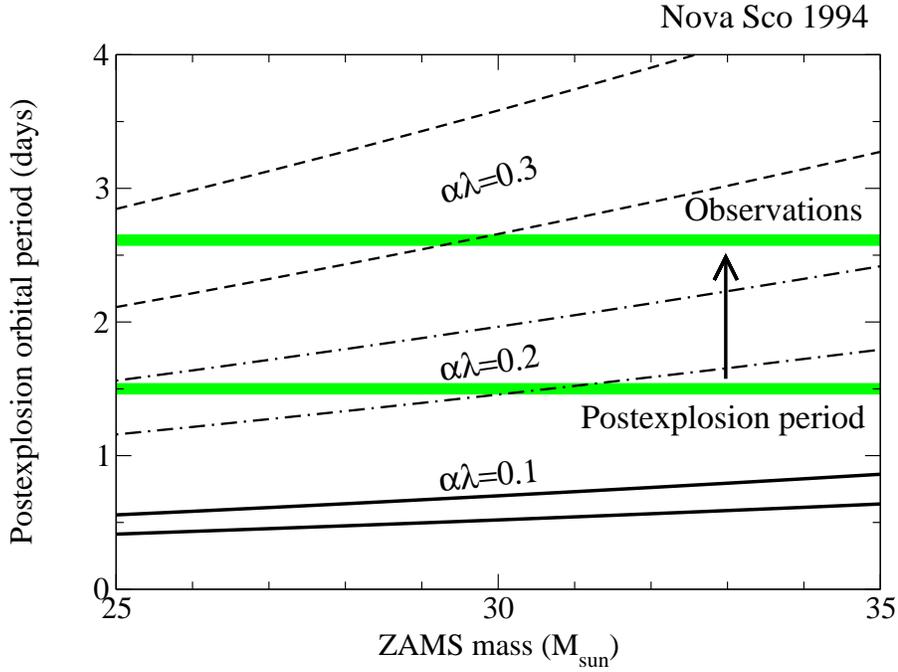,height=4.0in}}
\caption{Postexplosion orbital periods of Nova Sco
after BH formation for various common envelope efficiencies.
For our calculation, we use a post-explosion black hole mass 
$M_{BH}\sim 4.94 \msun$ and a
companion mass $M_d\sim 1.91 \msun$.
The x-axis is the ZAMS mass of black hole progenitor. 
The width of each band is determined by the initial band of 
possible Case C mass transfer separations given in Fig.~\ref{FIG4}. 
A post-explosion period of about 1.5\,day is 
consistent with $\lambda\alpha_{ce}\sim 0.2$.
By conservative mass transfer of $0.46\msun$ from the companion star
to the black hole, the orbital period can be evolved to
the currently observed orbital period. 
}
\label{FIGSco}
\end{figure}

Reconstruction of the AML binaries is more complicated, because they have 
lost angular momentum through magnetic braking and gravitational waves,
so that their present positions as plotted in Fig.~\ref{FIG1} are not
those at pre-explosion time. As with the evolved companions, matter will
have been accreted onto the black hole, so the black hole masses will
be somewhat greater than just following the explosion.
As noted earlier, the binaries with less massive companions with
separation $a_f$ at the end of common envelope evolution overfill their
Roche Lobes. The outer part of the companion, down to the Roche Lobe
$R_L$ is transferred onto the He star. This mass transfer widens the
orbit to $R_L$, possibly overshooting. Unless much mass is lost in
the explosion when the black hole is formed, the Roche lobe radius
is unchanged by the formation of the black hole, and corresponds to
line III in Fig.~\ref{FIG2}.

\begin{figure}
\centerline{\epsfig{file=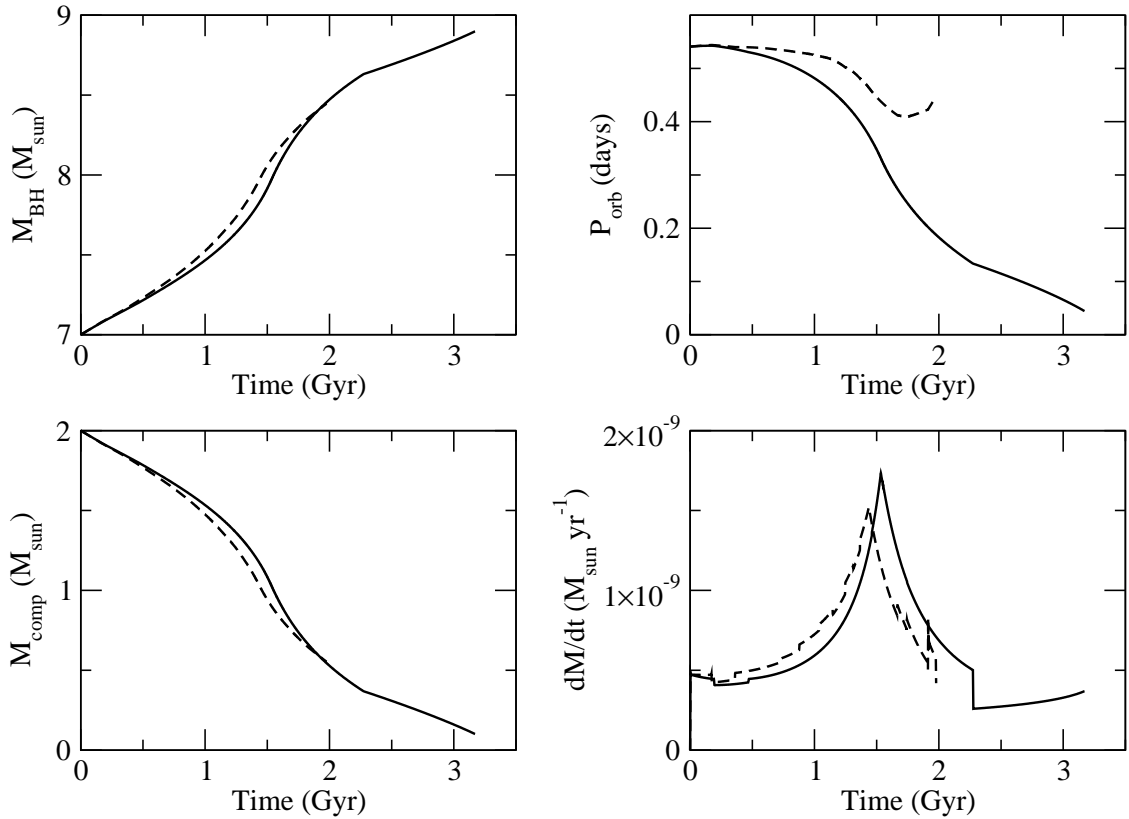,height=5.5in}}
\caption{Evolution of a binary with $7\msun$ black hole 
and $2\msun$ companion for the initial period of 0.54 day.
The solid line marks the evolution in case the companion star 
adjust itself as it loses mass;
the dashed line traces the evolution in case the mass loss does not
affect the internal time scale of the companion star, so that it
follows the same time evolution as an undisturbed $2\msun$ star.
}
\label{FIG7}
\end{figure}

\begin{figure}
\centerline{\epsfig{file=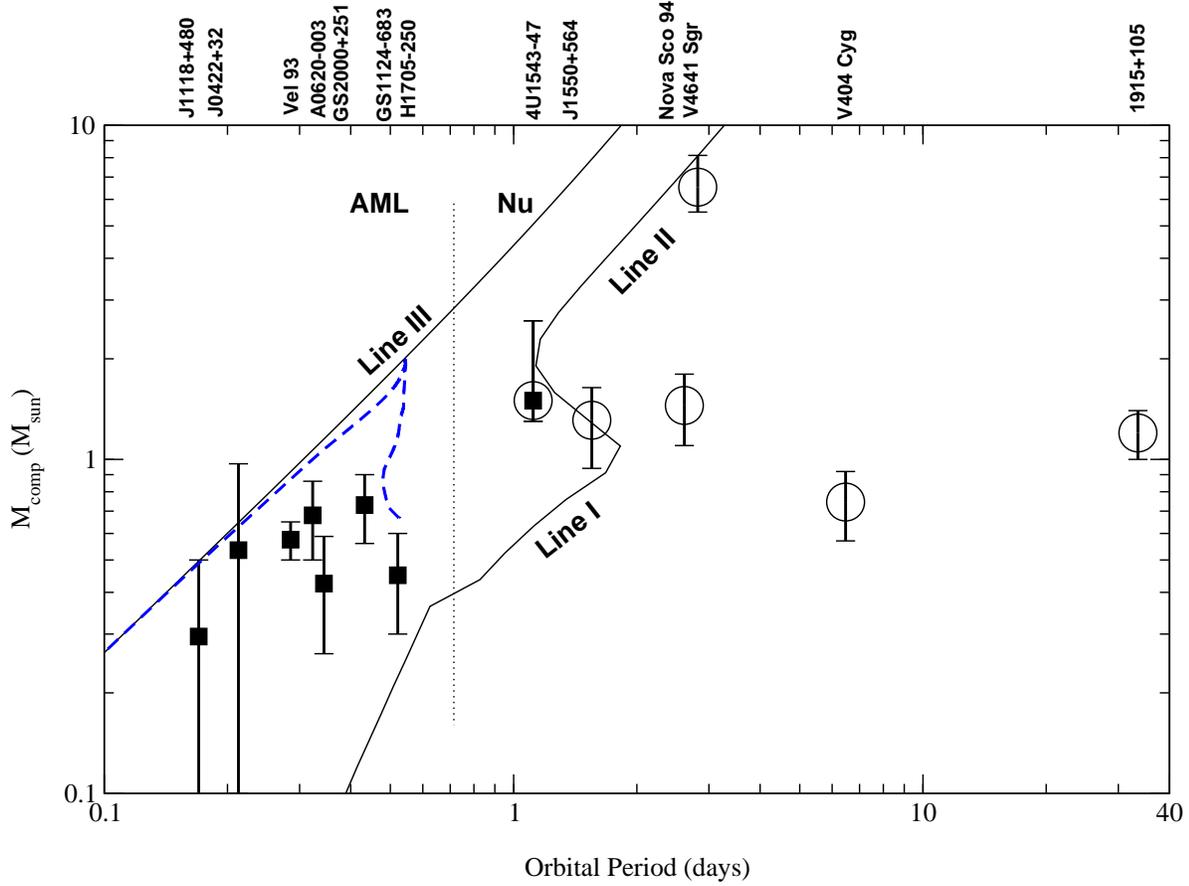,height=5in}}
\caption{Evolutionary tracks of a binary with $9\msun$ black hole 
and $2\msun$ companion for the initial period of 0.54 day.
The two evolution possibilities are as in Fig.~\ref{FIG7}.
Left (right) dashed line corresponds to the solid (dashed)
lines in Fig.~\ref{FIG7}.
}
\label{FIG8}
\end{figure}

\begin{figure}
\centerline{\epsfig{file=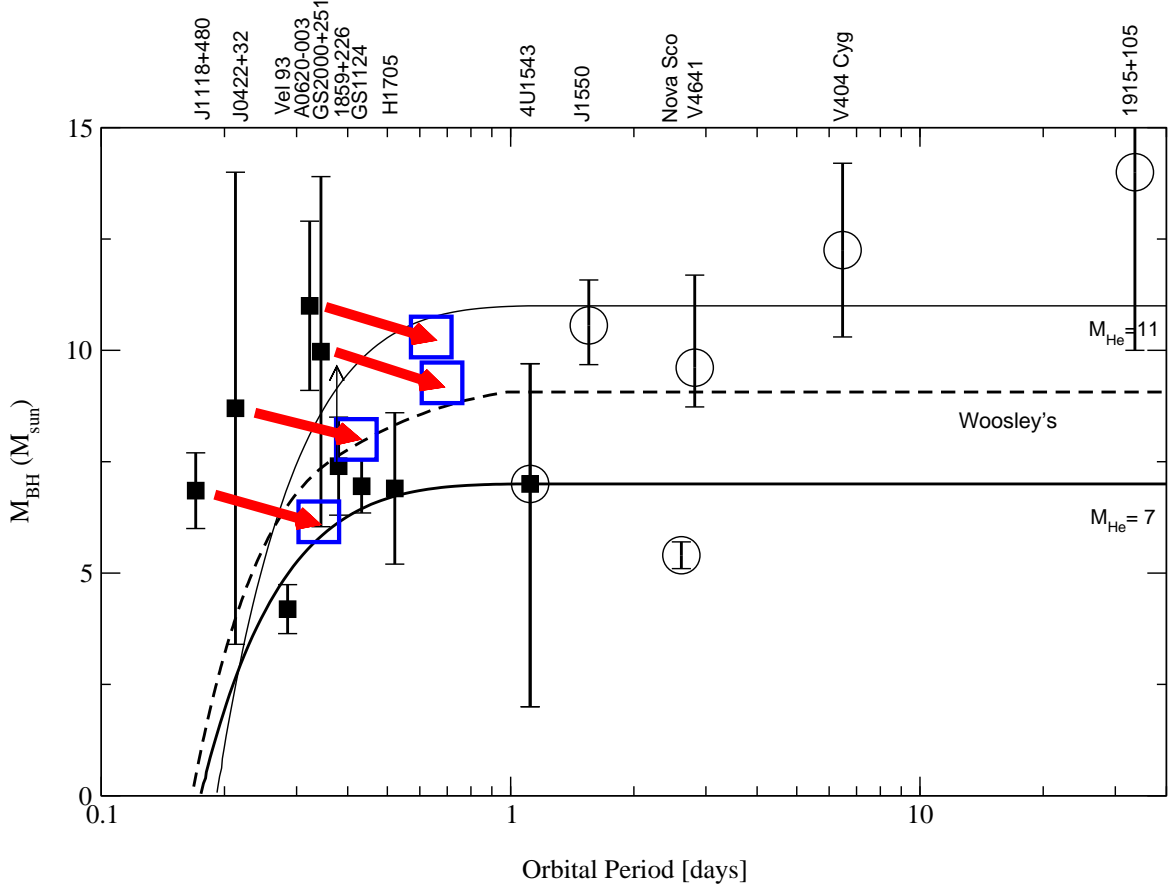,height=5in}}
\caption{Present orbital period vs.\ black hole masses 
of SXTs.  The deviations from the theoretical
curves are substantial due to the post-explosion
evolution of the binaries. The arrows on the AML systems point to an
approximate post-explosion location if the donor mass was initially 
1.5$\msun$, and 0.7$\msun$ has now been transferred to the black hole.
The solid lines 
indicate the possible ranges of black hole masses with 
polytropic index $n=3$ (radiative), for given
pre-explosion spin periods which are assumed the same as the pre-explosion
orbital period. Here we used $R_{He}=0.22 (M_{He}/\msun)^{0.6} \rsun$. 
For comparison, the results with a ``scaled" He core
($9.15\msun$) of Woosley's $25\msun$ star at the beginning of $^{12}$C
burning with $T_c= 5\times 10^8 K$, appropriate for Case C mass transfer, are
plotted as a dashed line (Woosley 2001). 
In this plot, we scaled the radius of Woosley's
core, $R_{Woosley}\sim 3\times 10$\,cm, by a factor 2. As can be seen, the
AML systems can plausible originate from systems within the curves, and
thus are consistent with our theory. However, since they could have originated
anywhere between the open square and their current location, they do not
strongly test the theory.
}
\label{FIG10}
\end{figure}

Brown, Lee, \& Tauris (2001c) explored the evolution of ZAMS
$1.25\msun$ stars under magnetic braking, gravitational waves and
mass transfer to the black hole. We adapt the same methods to make
a more detailed study of the AML.
First of all we construct (Fig.~\ref{FIG2}) the lower limit on the
companion mass for evolution in a Hubble time, giving the dashed line
there. All binaries  with companions in main sequence at the beginning
of mass transfer must lie between the dashed line and line III
in that figure.  The fact that the AMLs tend to lie below the dashed
line implies both mass loss from the companion and accretion onto the
black hole. Therefore, all these systems have shrunk their orbits and
increased their black-hole mass since the formation of the black hole,
by amounts that cannot be determined well.
In Fig.~\ref{FIG10}, however, we show where the four shortest period
AMLs would have come from, had they lost 0.7~$\msun$ from an
initial 1.5~$\msun$.
{}From our earlier discussion about the $a_f$ following common envelope
evolution we saw that binaries with companions which stayed in main
sequence were favored to come from companion masses less than $2\msun$,
and from Fig.~\ref{FIG2}, we see that they would chiefly have
companion ZAMS mass greater than $\sim 1\msun$, so that most of them would
initially have periods of 0.4--0.7 days (which follows from the separations
obtained from our eq.~(\ref{eqaf})).
In trying to understand the detailed evolution of the AML we begin
from a binary with a $2\msun$ companion which just fills its Roche Lobe
following common envelope evolution. We then follow its evolution under
the two assumptions made in Brown, Lee, \& Tauris (2001c): (1)
That its time of evolution is always given by its initial $2\msun$ mass,
i.e., ignoring effects of mass loss on the internal evolution
time (dashed lines in Fig.~\ref{FIG7}
and right dashed line in Fig.~\ref{FIG8}) 
(2) That the evolution of the star proceeds according
to its adjusted mass (solid lines in Fig.~\ref{FIG7} and left
dashed line in Fig.~\ref{FIG8}).
Since mass loss drives the companion out of thermal equilibrium, these
two extremes bracket the outcome of a full stellar model calculation.

In summary, the AML systems have had the information on their
post-explosion parameters partly erased by subsequent evolution, in a 
manner that we cannot undo. Therefore, they can only provide a crude
consistency check on the mass-period relation for black holes in SXTs,
rather than provide precise constraints.

\section{Angular momentum and its consequences for the mass and spin of
         the black hole\label{rotate}}

It is, in general, a difficult and unsolved problem to calculate the
angular momentum of a stellar core at any given time. Even if we make the
usual assumption that the rotation is initially solid-body, and not very
far away from the maximal stable rotation frequency, the viscous coupling
between the various layers of the star as it evolves is poorly known, and
thus it is hard to be very quantitative. The general trend, however, is that
the core will shrink and the envelope expand. In absence of viscous coupling,
every mass element retains its angular momentum, and hence the core spins
up as the envelope spins down, setting up a strong gradient in rotation
frequency between the core and the envelope. Viscosity will then act to reduce
this gradient, transporting angular momentum from the core to the envelope,
but the efficiency of this process is very uncertain
(Spruit \& Phinney 1998, Livio \& Pringle 1998).

As we noted above (section 1), in our scenario, a number of effects 
will increase the angular momentum of
the core relative to a similar core of a single star: (1)
during spiral-in, the matter somewhat inside the orbit of the secondary is
spun up by tidal torques (Rasio \& Livio 1996); (2) the removal of the
envelope halts the viscous slowdown of the core by friction with the envelope;
(3) during the post-spiral-in evolution, tidal coupling will tend to spin
the helium star up even closer to the orbital period than was achieved by
the first effect. This will not be a very strong effect because the duration
of this phase is short, but it will affect the outer parts of the
helium star somewhat, and this
is the most important part (see below).

The net result of all these effects will be that the helium star will spin
fairly rapidly, especially its envelope. The core is not so crucial to our
argument about the fraction of the star that can fall into the black hole,
since the few solar masses in it will not be centrifugally supported even
in quite short orbits. For the purpose of a definite calculation, we
therefore make the following assumptions: (1) the helium star co-rotates 
with the orbit before explosion and is in solid-body rotation; 
(2) the mass distribution of the helium
star with radius is given by a fully radiative zero-age helium main sequence
star. This latter approximation is, of course, not extremely good. However,
what counts is the angular momentum as a function of mass, so the fact that
the mass distribution has changed from helium ZAMS to explosion would be
entirely inconsequential if no redistribution of angular momentum had taken
place in the interim. As we saw above, any redistribution of angular momentum
would take the form of angular momentum transport toward the outer layers.
This means that relative to our ideal calculations below, a better calculation
would find more angular momentum in the outer layers, and therefore
somewhat smaller black hole masses than the ones we calculate.

We now investigate how much mass will be prevented from falling
into the black hole by the angular
momentum of the He star, under the above assumptions of solid-body rotation
with a period equal to that of the binary.
If we assume that angular momentum is conserved during the
collapse, we can get the cylindrical radius $R_c$ within which matter is
not centrifugally prevented from falling into the black hole:
   \be
   R_c^2\Omega =  \tilde l(\hat a) \frac{G M_c}{c} 
   \ee
where 
$\tilde l(\hat a)$ is the dimensionless
specific angular momentum of the marginally bound
orbit for a given Kerr parameter $\hat a$, and
$M_c$ is the total mass inside the cylinder of radius $R_c$.
The Kerr parameter becomes
  \be
 \hat a= \frac{I_c\Omega}{G M_c^2/c} = k^2 \tilde l(\hat a)
  \ee
where $I_c$ is the total moment of inertia inside the cylinder of radius $R_c$,
$I_c = k^2 M_c R_c^2$.
$M_c$ gives an estimate of the final black hole mass. Combining these
relations with a profile of angular momentum and mass versus radius 
using the assumptions listed above, we can calculate the expected black hole
mass and Kerr parameter as a function of SXT period before explosion.

In Fig.~\ref{FIG10} we show the predicted relation between orbital period
and black-hole mass for different helium star masses in our model. We
compare these with the {\it present\/} properties of all SXTs for which
the required parameters are known. The properties are consistent with 
the theoretical relations, but do not confirm it very strongly due to the
evolutionary changes discussed in section~\ref{preexp}. 
Specifically, the AML systems lie
above and to the left of the curves, because their orbits shrunk and their
black holes accreted mass since the formation of the black hole.
However, as we saw in section~\ref{preexp.aml}, plausible amounts of
conservative mass transfer since the explosion would place the systems
among the theoretical post-explosion curves (indicated by the open squares
and arrows).

To test the theory more strongly, we show in Fig.~\ref{FIG11} only those
systems for which the pre-explosion properties could be reconstructed
(section~\ref{preexp}).  We compare the observed
points with ideal polytropic helium stars of 7\,$M_\odot$ and 11\,$M_\odot$,
and with a full model calculation obtained from Woosley (2001). 
By coincidence, the
curves converge near the region of the shortest-period observed systems, so that
the uncertainty in helium star mass is not of great importance to the outcome.
A helium star mass in the lower end of the range (7--9\,$M_\odot$) may be
somewhat preferred for these systems. For periods above 1 day, angular-momentum
support is not important, and the 
mass of the final black hole will be very close to that of the helium star,
and thus varies somewhat from system to system. As we can see, the 
reconstructed pre-explosion properties lie much closer to the theoretical
predictions.

As a corollary, we find that systems with very large velocities, like Nova 
Sco, will be rare: at the shortest pre-explosion orbits, where much mass
is ejected, the companion mass tends to be small. Then the center of mass
of the binary is close to that of the helium star, which strongly limits the
systemic velocity induced by the mass loss. On the other hand, for the
widest systems, where the companion tends to be massive enough to allow a
significant systemic velocity induced by mass loss, the mass loss itself
becomes too small to induce much of a systemic velocity.

An important result for our proposed relation between SXTs and hypernovae
and gamma-ray bursts is shown in Fig.~\ref{FIG12}. This figure shows the 
expected Kerr parameter of the black hole formed in our model. We see that
for the short-period systems, this Kerr parameter is very large, 0.7--0.9.
This means that we are justified in adding only the mass that immediately
falls in to the black hole, because as soon as the rapidly rotating black hole
is formed, it will drive a very large energy flux in the manner described
by Brown et al.(2000). 
This both causes a GRB and expels the leftover stellar envelope.
The systems with longer orbital periods do not give rise to black holes
with large Kerr parameters, and thus are presumably not the sites of GRBs.

\section{Conclusions\label{concl}}

We have shown that there is an observed correlation between orbital period
and black-hole mass in soft X-ray transients. We have modeled this
correlation as resulting from the spin of the helium star progenitor of
the black hole: if the pre-explosion orbit has a short period, the helium star
spins rapidly. This means that some part of its outer envelope is centrifugally
prevented from falling into the black hole that forms at the core. This
material is then expelled swiftly, leading to a black hole mass much less
than the helium star mass.  As the orbital
period is lengthened, the centrifugal support wanes, leading to a more
massive black hole.  The reason for swift expulsion of material held up
by a centrifugal barrier is the fact that black holes formed in our
scenario naturally have high Kerr parameters (Fig.~\ref{FIG12}). This implies
that they input very high energy fluxes into their surrounding medium via
the Blandford-Znajek mechanism, and thus power both a gamma-ray burst and
the expulsion of the material that does not immediately fall in.

\begin{figure}
\centerline{\epsfig{file=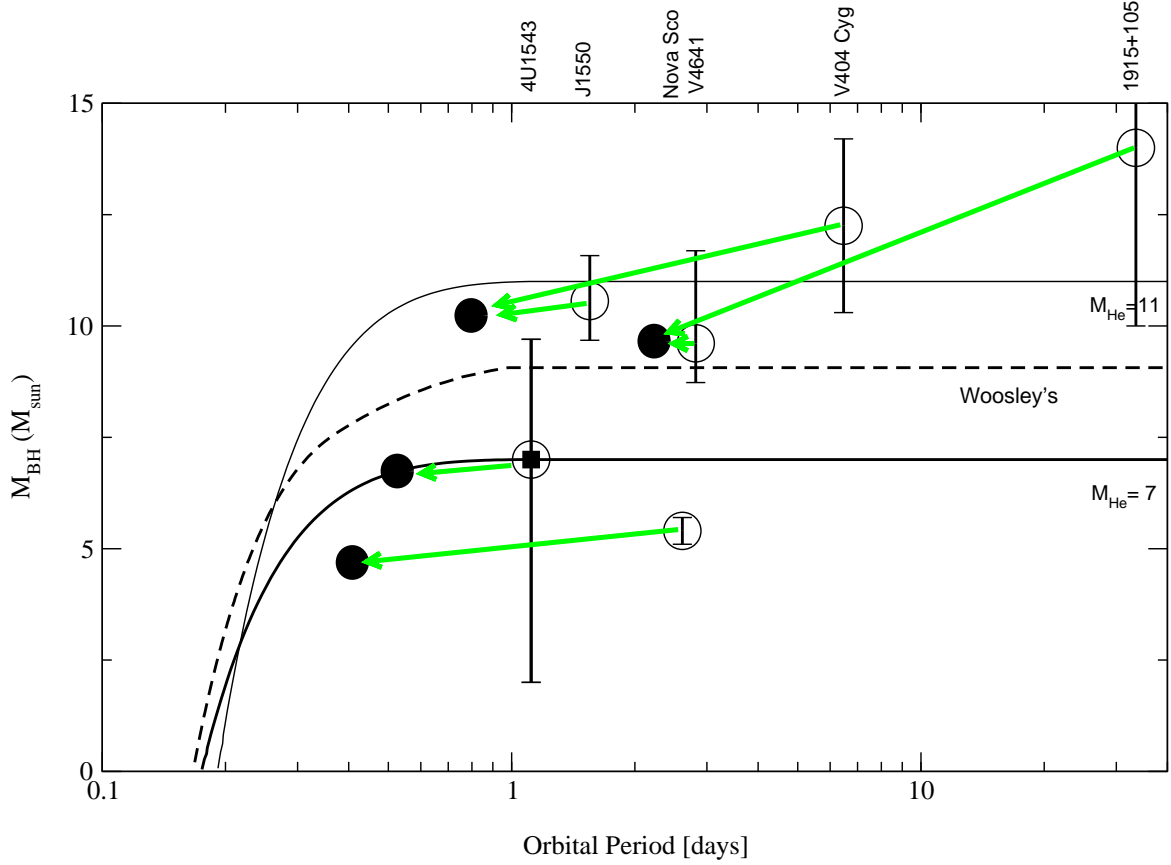,height=5in}}
\caption{Reconstructed pre-explosion orbital period vs.\ black hole masses 
of SXTs with evolved companions.
The reconstructed pre-explosion orbital periods and
black hole  masses are marked by filled circles, and
the current locations of binaries with evolved companions are marked by 
open circles.  
}
\label{FIG11}
\end{figure}

\begin{figure}
\centerline{\epsfig{file=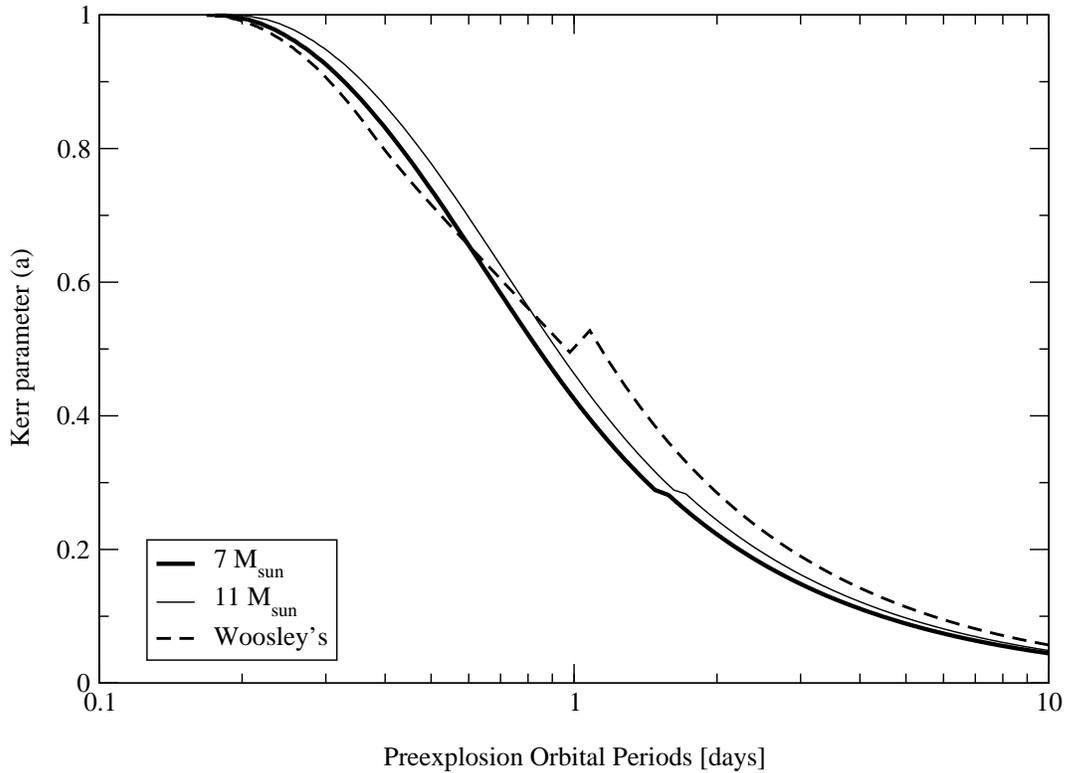,height=5in}}
\caption{The Kerr parameter of the black hole resulting from the collapse
of a helium star synchronous with the orbit, as a function of orbital period.
The conditions are the same as in Fig.~\ref{FIG10}, as is the meaning of the
three curves. 
Woosley's helium core is of mass $9.15\msun$ from a ZAMS $25\msun$ star
(Woosley 2001).
Note that the result depends very little on the mass of the
helium star, or on whether we use a simple polytrope or a more sophisticated
model. The plot illustrates that rapidly rotating black holes needed for
powering GRBs originate only from originally short-period SXTs.
\label{FIG12}
}
\end{figure}

However, because the correlation is induced between
the orbital period before explosion and the black-hole mass, its 
manifestation in the observed correlation between BH mass and present
orbital period is weakened due to
post-explosion evolution of the binaries. We therefore considered the
evolution in some detail, and for a subset of the systems were able to 
reconstruct the pre-explosion orbital periods.
The correlation between pre-explosion period and black hole mass 
(Fig.~\ref{FIG11}) is in much better agreement with our model
than the original one between present
period and black hole mass (Fig.~\ref{FIG1}). We developed a quantitative
model for the relation between period and mass, and showed that it fits the
subset of reconstructible SXT orbits.

Nova Scorpii stands out as the most extreme case of mass loss, nearly
half of the total system mass, and, therefore, a great widening
in the orbit which gets its period well beyond the gap between shrinking
and expanding orbits.
{}From Fig.~\ref{FIG11} we see that its black hole mass is far below the
polytropic line for its $M_{\rm He}=11\msun$ progenitor. We believe
that in the case of this binary a short central engine time of
several seconds was able to furnish angular momentum and energy
to the disk quickly enough to stop the infall of some of the
interior matter not initially supported by centrifugal force;
i.e., the angular momentum was provided in less than a dynamical time.
In other words, the Blandford-Znajek mechanism that drives
the GRB not only expelled
the matter initially supported for a viscous time by angular momentum,
but actually stopped the infall within a dynamical time.

Since we can also compute the Kerr parameters of the black holes formed via
our model, we find that the short-period systems should have formed black 
holes with Kerr parameters in the range 0.7--0.9. This makes them prime
candidates for hypernovae and gamma-ray bursts, and thus provides further
support for our earlier study in which we posited that SXTs with black-hole
primaries are the descendants of gamma-ray bursts. We can now also refine this
statement: SXTs {\it with short orbital periods\/} before the formation of the
black hole have given rise to a GRB in the past.

\section*{Acknowledgments}

We would like to thank J. Orosz  and S. Woosley
for useful advice and information.
Hans Bethe critically read the manuscript and offered helpful
suggestions.
We are very grateful to the referee for a number of helpful suggestions,
which enabled us improve the manuscript. In particular he brought
to our attention the Greiner et al. (2001) observations of 1915$+$105,
which we could evolve as V4641 Sgr viewed at a later time.
CHL was in part supported by the BK21 project of the Korea Ministry of
Education.
We were in part supported by the U.S. Department of Energy under grant
DE-FG02-88ER40388. RAMJW also acknowledges partial support from NASA
under grant NAG510772.

\bibliographystyle{astron}

\newpage
\begin{table}
\footnotesize
\def\arraystretch{0.9}
\newcommand{\ti}[1]{{\footnotesize #1}}
\begin{tabular}{@{}llccccl@{}}\hline
                    &                   &  compan.         & $P_{orb}$             & $f(M_{X})$ & $M_{opt}$              &      \\ 
X-ray               & other             &  type            & (day)                 &  ($\msun$) & ($\msun$)              & Ref. \\
names               & name(s)           &  $K_{opt}$       &      $d$              &  i         & $M_{BH}$               & \\ 
                    &                   &  (km s$^{-1}$)   & (\kpc)                &  (degree)  & ($\msun$)              & \\
\hline
XTE J1118$+$480     &\ti{KV Ursae Majoris}    & K7V-M0V    & 0.169930(4)           & 6.1(3)     &  0.09--0.5             & 1,2) \\
                    &                         & 701(10)    & 1.9(4)                & 81(2)      &  6.0--7.7              & \\ \hline
XN Per 92           & \ti{V518 Persei}        &  M0 V      & 0.2127(7)             & 1.15--1.27 & 0.10-0.97              & 3)\\
\ti{GRO\,J0422$+$32}&                         &  380.6(65) &                       & 28--45     & 3.4--14.0              & \\ \hline
XN Vel 93           & \ti{MM Velorum}         & K6-M0      & 0.2852                & 3.05--3.29 & 0.50--0.65             & 4) \\
                    &                         &  475.4(59) &                       & $\sim$ 78  &  3.64--4.74            & \\ \hline
XN Mon 75           &\ti{V616\,Monocerotis}   &  K4 V      & 0.3230                & 2.83-2.99  & 0.68(18)               & 3,5)\\
\ti{A\,0620$-$003}  &\ti{N Mon 1917}          & 433(3)     &  1.164(114)           & 40.75(300) & 11.0(19)               & \\ \hline
XN Vul 88           &\ti{QZ\,Vulpeculae}      &  K5 V      & 0.3441                &  5.01(12)  & 0.26--0.59             & 3,6) \\
\ti{GS\,2000$+$251} &                         &   520(16)  & 2                     &  47--75    &  6.04-13.9             & \\ \hline
XTE 1859$+$226      & \ti{V406 Vulpeculae}    &            &  0.380(3)             & 7.4(11)    &                        & 7)  \\
                    &                         & 570(27)    &                       &            &                        & \\ \hline
XN Mus 91           & \ti{GU Muscae}          &  K4 V      & 0.4326                & 2.86--3.16 & 0.56--0.90             & 3,8) \\
\ti{GS\,1124$-$683} &                         &  406(7)    & 3.0                   & 54.0(15)   & 6.95(6)                & \\ \hline
XN Oph 77           &\ti{V2107\,Ophiuchi}     &  K3 V      & 0.5213                & 4.44--4.86 &  0.3--0.6              & 3)\\
\ti{H\,1705$-$250}  &                         &   420(30)  &  5.5                  &   60--80   &  5.2--8.6              & \\ \hline
XN                  &\ti{MX 1543-475}         &  A2 V      & 1.1164                & 0.252(11)  &  1.3--2.6              & 9,10) \\
\ti{4U 1543$-$47}   & \ti{IL Lupi}            &  129.6(18) &  9.1(11)              & $\sim$ 22  &  2.0--9.7              & \\ \hline
XTE J1550$-$564     & \ti{V381 Normae}        & G8IV-K4III &  1.552(10)            & 6.86(71)   & 1.31$^{+0.33}_{-0.37}$ & 11)\\
                    &                         & 349(12)    &  4.7-5.9 (?)          & 70.8--75.4 & 10.56$^{+1.02}_{-0.88}$& \\ \hline
XN Sco 94           & \ti{V1033 Scorpii}      &  F6III     & 2.6127(8)             & 2.64--2.82 & 1.1--1.8               & 3,12) \\
\ti{GRO\,J1655$-$40}&                         &  227(2)    & 3.2                   &   67--71   & 5.1--5.7               & \\ \hline
V4641 Sagittarii    & \ti{SAX J1819.3$-$2525} &  B9III     &  2.817                & 2.74(12)   & 6.53$^{+1.6}_{-1.03}$  & 13) \\
\ti{XTE J1819$-$254}&                         &  211.0(31) & 9.59$^{+2.72}_{-2.19}$&            & 9.61$^{+2.08}_{-0.88}$ & \\ \hline
Cyg\,X-1            &\ti{V1357\,Cyg}          &  O9.7Iab   &  5.5996               & 0.25(1)    & $\sim$ 17.8            & 14) \\
\ti{1956$+$350}     &\ti{HDE\,226868}         & 74.7(10)   &  2.5                  &            & $\sim$ 10.1            & \\ \hline
XN Cyg 89           &\ti{V404\,Cygni}         &  K0 IV     & 6.4714                & 6.02--6.12 & 0.57--0.92             & 3,15,16) \\
\ti{GS\,2023$+$338} &\ti{N Cyg 1938, 1959}    & 208.5(7)   & 2.2-3.7               &   52--60   & 10.3--14.2             & \\ \hline
GRS 1915$+$105      & \ti{V1487 Aquilae}      & K-MIII     &  33.5(15)             & 9.5(30)    &  1.2(2)                & 17) \\
                    &                         & 140(15)    &  12.1(8)              & 70(2)      &  14(4)                 & \\ \hline
\end{tabular}
\caption{Parameters of black hole binaries in our Galaxy with measured mass functions.
Binaries are listed in order of increasing orbital period.
All systems except Cyg~X-1 (steady X-ray source) are soft X-ray transients.
         XN indicates X-ray Nova.
Earlier observations (Greene et al. 2001) gave
the black hole mass in Nova Scorpii as $6.3\pm 0.5\msun$
New analyses of the light curve by Beer \& Podsiadlowski (2001)
give a somewhat smaller mass $5.4\pm 0.3 \msun$ and $1.45\pm 0.35\msun$
for the companion. 
References: 
$^{1)}$ McClintock et al. 2001,
$^{2)}$ Wagner et al. 2001,
$^{3)}$ Bailyn et al. 1998,
$^{4)}$ Filippenko et al. 1999,
$^{5)}$ Gelino et al. 2001a,
$^{6)}$ Harlaftis et al. 1996,
$^{7)}$ Filippenko \& Chornock 2001. 
$^{8)}$ Gelino et al. 2001b,
$^{9)}$ Orosz et al. 1998,
$^{10)}$ Orosz  2002, 
$^{11)}$ Orosz et al. 2002, 
$^{12)}$ Beer \& Podsiadlowski 2001,
$^{13)}$ Orosz et al. 2001,
$^{14)}$ Herrero et al. 1995,
$^{15)}$ Shahbaz et al. 1994,
$^{16)}$ Shahbaz et al. 1996,
$^{17)}$ Greiner et al. 2001. 
}
\label{tab1}
\end{table}

\begin{table}
\begin{center}
\begin{tabular}{lcc}
\hline
Parameter                    & Nova Scorpii    & V4641 Sgr \\
\hline
Orbital period (days)        &  2.623          & 2.817 \\
Black hole mass ($\msun$)    &  $5.4\pm 0.3$   & $9.61^{+2.08}_{-0.88}$ \\
Companion mass  ($\msun$)    &  $1.45\pm 0.35$ & $6.53^{+1.6}_{-1.03}$ \\
Total mass ($\msun$)         &        6.85     & $16.19^{+3.58}_{-1.94} $ \\
Mass ratio                   &        0.27     & $1.50\pm 0.13$ \\
Orbital separation ($\rsun$) &       15.2      & $21.33^{+1.25}_{-1.02}$   \\
Companion radius ($\rsun$)   &        4.15     & $7.47^{+0.53}_{-0.47}$ \\
Distance (kpc)               &        3.2      & $9.59^{+2.72}_{-2.19}$ \\
\hline
\end{tabular}
\end{center}
\caption{Parameters for Nova Scorpii (Beer \& Podsiadlowski 2001)
 and V4641 Sgr (Orosz et al. 2001).}
\label{tab2}
\end{table}

\newpage

\end{document}